\documentclass[
    ,final            
  ]
  {aipproc}

\layoutstyle{8x11double}

\begin{document}

\title{The GBT350 Survey of the Northern Galactic Plane for
Radio Pulsars and Transients}


\author{J.~W.~T. Hessels}{
  address={Astronomical Institute ``Anton Pannekoek'', University of
Amsterdam, Kruislaan 403, 1098 SJ Amsterdam, The Netherlands}}

\author{S.~M. Ransom}{
  address={National Radio Astronomy Observatory, 520 Edgemont Road,
Charlottesville, VA 22903, U.S.A.}}

\author{V.~M. Kaspi}{
  address={Department of Physics, McGill University, Montreal, QC H3A 2T8,
Canada}}

\author{M.~S.~E. Roberts}{
  address={Eureka Scientific, Inc., Oakland, CA 94602-3017, U.S.A.}}

\author{D.~J. Champion}{
  address={Department of Physics, McGill University, Montreal, QC H3A 2T8,
Canada}}

\author{B.~W. Stappers}{
  address={Jodrell Bank Centre for Astrophysics, University of Manchester,
   Manchester, M13 9PL, UK}}
  
\begin{abstract}

Using the Green Bank Telescope (GBT) and Pulsar Spigot at 350\,MHz, we have
surveyed the Northern Galactic Plane for pulsars and radio transients.  This
survey covers roughly 1000 square degrees of sky within $75^{\circ} < l <
165^{\circ}$ and $|b| < 5.5^{\circ}$, a region of the Galactic Plane
inaccessible to both the Parkes and Arecibo multibeam surveys.  The large
gain of the GBT along with the high time and frequency resolution provided
by the Spigot make this survey more sensitive by factors of about 4 to slow
pulsars and more than 10 to millisecond pulsars (MSPs), compared with
previous surveys of this area.  In a preliminary, reduced-resolution search
of all the survey data, we have discovered 33 new pulsars, almost doubling
the number of known pulsars in this part of the Galaxy.  While most of these
sources were discovered by normal periodicity searches, 5 of these sources
were first identified through single, dispersed bursts.  We discuss the
interesting properties of some of these new sources.  Data processing using
the data's full-resolution is ongoing, with the goal of uncovering MSPs
missed by our first, coarse round of processing.

\end{abstract}


\maketitle


\section{Introduction}

Here we define the ``Northern Sky'' as the sky above the northern Arecibo
declination limit of $38^{\circ}$.  Previous pulsar surveys of the Northern
Sky between $300-500$\,MHz include an all-sky survey above $\delta >
38^{\circ}$ using the Jodrell Bank telescope at 428\,MHz, which, largely due
to RFI, discovered only one pulsar \citep{nll+95}; a survey at Green Bank
using the 140-ft (43-m) equatorially mounted telescope at 370\,MHz, which
found eight new pulsars \citep{snt97}; and a survey using the Green Bank
300-ft (91-m) transit telescope at 390\,MHz, which provided partial coverage
of the Northern Sky and discovered 20 pulsars \citep{sstd86}. 

The GBT's large gain (2\,K/Jy) and maneuverability make it the most
sensitive radio telescope on Earth over a large portion of the sky.
Furthermore, the primary beam width of the GBT at 350\,MHz is $\sim
0.6^{\circ}$, large enough to permit very efficient large-scale,
single-pixel surveys of the sky.  This motivated us to commence a 350-MHz
survey, which we will refer to as the GBT350 survey, of the Northern
Galactic Plane, corresponding to Galactic longitudes
$75^{\circ} < l < 165^{\circ}$ and out to latitudes $|b| < 5.5^{\circ}$.
Here we present a short summary of the survey design and analysis
techniques, and then discuss some of the new sources.

\section{Observations \& Analysis}

\begin{figure}
\includegraphics[height=.35\textheight,angle=-90]{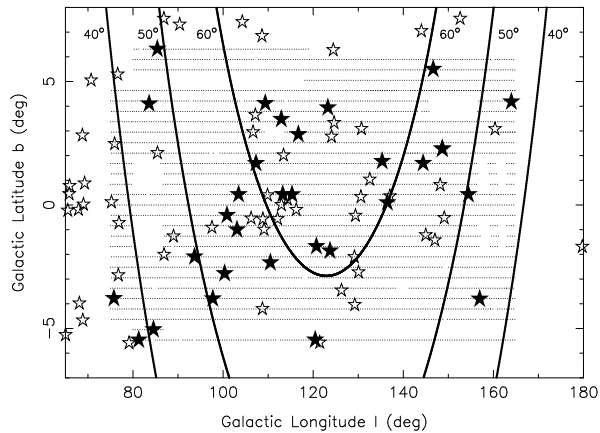}
\caption{The GBT350 survey area shown in Galactic coordinates.  Observed
pointing positions are marked with dots.  Empty and filled stars denote the
positions of known and newly found pulsars respectively.  Constant lines of
declination at 40$^{\circ}$, 50$^{\circ}$, and 60$^{\circ}$ are also shown.}
\label{coverage.fig}
\end{figure}

Data collection began in April 2005 and is now mostly complete.  A total of
$\sim 4000$ pointings were required to cover the survey area, which is shown
in Figure~\ref{coverage.fig}. Data were recorded at a centre frequency of
350\,MHz using the GBT Pulsar Spigot \citep{kel+05} in a 50\,MHz bandwidth
mode with 1024 lags and 81.92\,$\mu$s sampling.  With this spectral
resolution, there is a residual dispersive smearing of 100~(1000)\,$\mu$s at
DM = 10~(100)\,pc cm$^{-3}$.  Thus, these data provide good sensitivity to
even the fastest known MSPs, up to a DM of roughly 100\,pc cm$^{-3}$. The
integration time per pointing is 120\,s, resulting in a survey sensitivity
to normal, slow pulsars of roughly $S_{400}^{\rm min} = 2$\,mJy.  The GBT350
survey is a factor of $> 4$ times more sensitive to slow pulsars and $> 10$
times more sensitive to fast pulsars at moderate DMs than previous surveys
of the same area.  Furthermore, the current RFI environment at Green Bank in
this frequency range is remarkably good.  This allows us to use a very low
signal-to-noise threshold when investigating candidates.

We have analyzed all the survey pointings at reduced resolution in order to
first pick out the slower ($P_{\rm spin} > 50$\,ms) pulsars.  This
processing takes only 10\% the time of full resolution processing and is
capable of finding the vast majority of the potentially detectable sources.
We are currently re-processing the data at full resolution and expect that
this will reveal a few more sources, such as MSPs, whose properties make
them undetectable at lower resolution.  We have employed both standard
periodicity searches, summing up to 16 harmonics and including a narrow
search in acceleration space, as well as single pulse searches for bright,
dispersed bursts.  We are also considering searching the data for very slow
($P_{\rm spin} > 4$\,s) pulsars with a fast-folding algorithm, as well as
re-searching the data using a larger number of summed harmonics to increase
sensitivity to very short-duty cycle ($< 2$\%) pulsars.

\section{Results}

Thus far, we have discovered 33 pulsars, of which 5 were discovered through
their bright pulses in our single pulse search.  There are 39 previously
known pulsars in this part of the Galaxy and thus the GBT350 survey has
significantly increased the known pulsar population in the Nothern Galactic
Plane.  These new sources and their basic properties are summarized in
Table~\ref{pulsars.tab}.  We are timing all these new pulsars with GBT and
Westerbork and have timing solutions for 12 so far.  We now briefly discuss
the interesting properties of some of the new systems:

\begin{table}
\begin{tabular}{lccl}
\hline
  \tablehead{1}{l}{b}{Pulsar}
  & \tablehead{1}{c}{b}{Period}
  & \tablehead{1}{c}{b}{DM}
  & \tablehead{1}{c}{b}{Notes} \\
  \tablehead{1}{l}{b}{}
  & \tablehead{1}{c}{b}{(ms)}
  & \tablehead{1}{c}{b}{(pc cm$^{-3}$)}
  & \tablehead{1}{c}{b}{} \\
\hline
J0033+57       & 315    & 76  & \\
J0033+61       & 912    & 37  & \\
J0054+66       & 1390   & 15  & Nearby, RRAT\\
J0058+6125$^*$ & 637    & 129 & Extreme Nuller\\
J0240+62       & 592    & 4   & Nearby, Low-lum\\
J0243+6027$^*$ & 1473   & 141 & \\
J0341+5711$^*$ & 1888   & 100 & \\
J0408+55A      & 1837   & 55  & \\
J0408+55B      & 754    & 64  & \\
J0413+58       & 687    & 57  & \\
J0419+44       & 1241   & 71  & \\
J0426+4933$^*$ & 922    & 85  & \\
J0519+44       & 515    & 52  & \\
J2024+48       & 1262   & 99  & \\
J2029+45       & 1099   & 228 & \\
J2030+55       & 579    & 60  & \\
J2038+35       & 160    & 58  & \\
J2043+7045$^*$ & 588    & 57  & \\
J2102+38       & 1190   & 85  & \\
J2111+40       & 4061   & 120 & Slow rotator\\
J2138+4911$^*$ & 696    & 168 & \\
J2203+50       & 745    & 79  & \\
J2208+5500$^*$ & 933    & 105 & \\
J2213+53       & 751    & 161 & \\
J2217+5733$^*$ & 1057   & 130 & \\
J2222+5602$^*$ & 1336   & 168 & Periodic Nuller \\
J2238+6021$^*$ & 3070   & 182 & Slow rotator\\
J2244+63       & 461    & 92  & \\
J2315+58       & 1061   & 74  & \\
J2316+64       & 216    & 248 & \\
J2326+6141$^*$ & 790    & 33  & \\
J2343+6221$^*$ & 1799   & 117 & \\
J2352+65       & 1164   & 152 & \\
\hline
\end{tabular}
\caption{\footnotesize New Pulsars from the GBT350 Survey.  We are timing
all of these sources, and those marked with an asterisk already have precise
astrometric and spin parameters.  The source PSR~J2029+45 was also
discovered with Westerbork (Stappers \& Janssen, private communication) and
the sources PSRs~J2208+5500 and J2217+5733 with GMRT (McLaughlin, private
communication, also in this volume).} \label{pulsars.tab}
\end{table}

\noindent {\bf PSR~J0426+4933}: Westerbork observations of this pulsar at
1.4\,GHz have revealed two distinct components in its cumulative pulse
profile. Preliminary analysis indicates that the second, brighter, and
extremely narrow component is composed of individual bright pulses.  The
duty cycle of these individual pulses is only $\sim0.2$\%, roughly 2\,ms.
Deeper studies into the pulse energy distribution and single pulse
characteristic of PSR~J0426+4933 are planned.  Are these pulses ``giant
pulse like'', or is PSR~J0426+4933, like PSR~B0656+14 \citep{wwsr06}, a
pulsar that falls in between the groups of pulsars with `giant pulses'',
`giant micropulses'' \citep{jvkb01}, and the RRATs \citep{mll+06}?

\noindent {\bf PSR~J2222+5601}: This pulsar shows what looks like
quasi-periodic nulling, with roughly constant emission for $\sim 120$\,s
followed by $\sim 30$\,s gaps.  We are using the GBT to investigate how
strictly periodic these nulls are.  A few other examples of pulsars showing
periodic nulls have recently been identified (e.g. PSR~B1133+16, Herfindal
et al. in this volume). This emission behaviour is also reminiscent of that
recently identified in the so-called `sometimes a pulsar'' PSR~B1931+24
\citep{klo+06}, although the behaviour seen here is on a much shorter
timescale.

\noindent {\bf PSR~J0058+6125}: This pulsar is nulled during approximately
97\% of its pulse periods, placing it closer to the RRATs \citep{mll+06}
than the normal pulsar population in terms of its visibility.  This is a
considerably larger nulling fraction than those typically observed in nulling
pulsars \citep[consider for instance][]{wmj07}.  Because of the difficulty
in discovering such an intermittent pulsar, many similar systems are likely
to exist, but have gone undetected. Nulling of PSR~J0058+6125 has been
observed to last for up to about 20\,min (close to 2000 pulse periods) and,
intriguingly, the emission appears to come in windows of $100-200$\,s, in
which numerous, closely spaced pulses are seen. Nulling also occurs within
windows where the pulsar is visible.  Thus, it appears that this pulsar has
two nulling timescales: one on the order of the pulse period, the other on
the order of several minutes.


\noindent {\bf PSR~J0240+62}: This source, identified by its bright pulses
in our single pulse search, has a DM of only $\sim 4$\,pc cm$^{-3}$.  This
is extremely low (only 6 of the $\sim 1800$ known pulsars have a lower DM),
and suggests that this pulsar is very nearby.  Using the NE2001 electron
model of the Galaxy \cite{cl02}, the inferred distance is a mere 400\,pc,
making it one of the closest known pulsars to Earth.  Follow-up observations
of this source are needed to determine its radio flux.  If the source is
indeed as weak as it appears to be in its discovery observation, then its
proximity also indicates that it is a very low-luminosity source. Such
sources are important for mapping the low-luminosity end of the pulsar
luminosity distribution and the discovery of this source bodes well for the
discovery of more such sources with LOFAR \citep[see][for a description of
possible pulsar-related LOFAR science]{svk+07}.  Due to its proximity, we are
also planning X-ray observations of this source.

\begin{figure}
\includegraphics[height=.35\textheight,angle=-90]{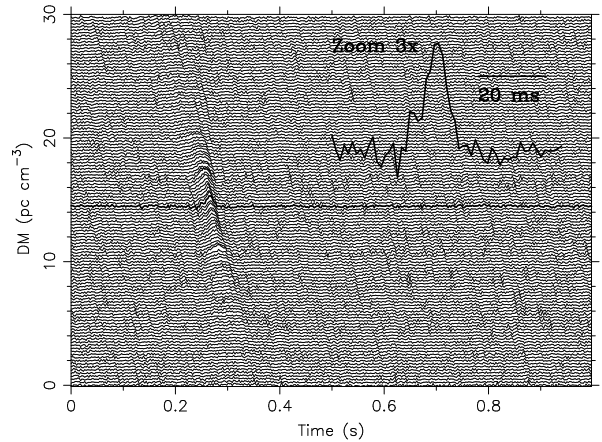}
\caption{{\it Main Panel:} A 1-s section of the discovery observation of
PSR~J0054+66.  A bright burst is shown as a function of DM
and time.  The signal peaks at 14.5\,pc cm$^{-3}$ and is smeared at DMs above
and below this. {\it Inset:} Zoom-in of the same burst, properly
dedispersed.  The burst has a least two components.}
\label{burst.fig}
\end{figure}

\noindent {\bf PSR~J0054+66}: This source was very recently discovered via
its occasional strong, dispersed bursts, one of which is shown in
Figure~\ref{burst.fig}. Though we have currently only observed 8 bursts from
this source, it appears that they are spaced in multiples of 1.3903\,s and
occur once every $\sim 100-1000$\,s.  This implies that PSR~J0054+66 is an
RRAT-like source, showing very sporadic radio pulsations.  It has a low DM
of 14.5\,pc cm$^{-3}$ and a corresponding DM distance of only 1\,kpc.
PSR~J0054+66's proximity bodes well for future studies in X-rays. What is
needed to better study this source in radio is longer continuous
observations to better characterize the distribution of the bursts in
intensity and time.  This will then determine how this intriguing new source
fits in with the known population of RRATs.

\begin{theacknowledgments}
J.W.T.H. thanks NSERC and the Canadian Space Agenncy for a postdoctoral
fellowship and supplement respectively.
\end{theacknowledgments}


\bibliographystyle{aipproc}   


\bibliography{Hessels_Jason_1}


\end{document}